\documentclass[fleqn,twoside]{article}
\usepackage{espcrc2}

\usepackage{graphicx}
\usepackage{epsfig}
\usepackage[figuresright]{rotating}

\newcommand{\AmS}{{\protect\the\textfont2
  A\kern-.1667em\lower.5ex\hbox{M}\kern-.125emS}}
\def\pme{$\pi^+ \rightarrow \mu^+ \rightarrow e^+$}
\def\pnn{$K \rightarrow\pi\nu\overline\nu$}

\def\kpnn{$K^+\rightarrow\pi^+\nu\overline\nu$}
\def\kpitwo{$K^+\rightarrow\pi^+\pi^0$}
\def\kmutwo{$K^+\rightarrow\mu^+\nu$}

\def\kpimue{$K^+ \rightarrow \pi^+ \mu^+ e^-$}

\def\kpimue{$K^+ \rightarrow \pi^+ \mu^+ e^-$}
\def\klpnn{$K_L\rightarrow\pi^0\nu\overline\nu$}
\def\klmue{$K_L \rightarrow \mu^{\pm} e^{\mp}$}
\def\kp{$K_{\pi2}$}
\def\km{$K_{\mu2}$}
\def\kmm{$K_{\mu m}$}
\def\klpitwo{$K_L \rightarrow \pi^0 \pi^0$}
\def\klethreeg{$K_L \rightarrow \pi^- e^+ \nu \gamma$}
\def\epistar{$E^{*}_{\pi^0}$}
\def\ediff{$|E^{*}_{\gamma 1} - E^{*}_{\gamma 2}|$}
\title{Plans for Kaon Physics at BNL}

\author{G. Redlinger\address{Physics Department, Brookhaven National
    Laboratory\\
    Upton, New York, USA 11973}
    \thanks{Invited talk presented
at HIF04: High Intensity Frontier Workshop, La Biodola, Isola d'Elba,
June 5-8, 2004}
}
       
\begin{document}

\begin{abstract}
I give an overview of current plans for kaon physics at BNL.  The
program is centered around the rare decay modes \kpnn~ and
\klpnn.
\vspace{1pc}
\end{abstract}

\maketitle

\section{INTRODUCTION}

There is a long history of kaon physics at the high intensity frontier
at the BNL AGS with results going back into the lore of high energy
physics. Interest in rare
kaon decays at BNL rose anew in the early 1980's with beam intensities
starting at $10\times10^{12}$ protons per spill; by the time the
program came to an end in 1998, beam intensities were regularly reaching
$60\times 10^{12}$ protons per spill with a 55\% duty cycle.
To cite a few ``flagship'' results, this era produced world-record
limits on
lepton flavor violating decays such as \klmue~\cite{kmue} and
\kpimue~\cite{kpimue}, improving limits by 6 and 3 orders of magnitude,
respectively.  Three orders of magnitude for the decay
\kpnn~ were covered, resulting in the observation of two events for this
mode \cite{E787pnn1}, at a level statistically consistent with
Standard Model expectations, but with a central value of the branching
ratio tantalizingly high by a factor of about two.

A new era began at BNL in 2000 with the start of the Relativistic Heavy Ion
Collider (RHIC) program.  The AGS is used as a heavy ion injector to
the RHIC ring, but can also be used to accelerate protons in between
RHIC fills, allowing a kaon program (for example) to operate
concurrently with RHIC at low
incremental cost.  This kaon program is centered on the rare decays
\kpnn~ and \klpnn~ with the corresponding experiments
E949 \cite{E949web}
and KOPIO \cite{KOPIOweb}.

\begin{figure}[htb]
\begin{center}
\epsfig{file=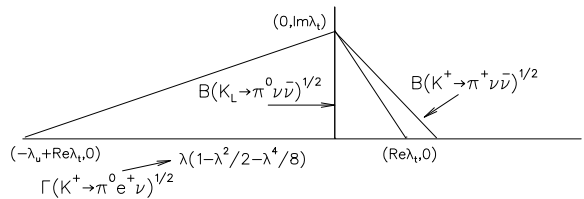,width=6pc,angle=90}
\caption{The unitarity triangle determined entirely from kaon decays.
  The triangle is not drawn to scale; the base is actually about 1000 times
  longer than the height. From \cite{Kettell}.}
\label{fig:kut}
\end{center}
\end{figure}

The literature on the decays \kpnn~ and \klpnn~ go back over 30
years \cite{Gaillard}.  For a recent perspective, 
a  good starting point is the review \cite{Buras04}.  The physics
interest in these decay modes
comes from the potential to completely determine the Unitarity
Triangle from kaon decays alone.  This is shown in Figure
\ref{fig:kut}, which shows the ``kaon'' unitarity relation
$V_{us}^*V_{ud} + V_{cs}^*V_{cd} + V_{ts}^*V_{td} = 0$ or
$\lambda_u + \lambda_c + \lambda_t = 0$ where $\lambda_i =
V_{is}^*V_{id}$.

An inconsistency between the unitarity relation in kaon decays
($s\rightarrow d$ transitions)
with that in B decays ($b\rightarrow d$ sector) would be a sign of
physics beyond the Standard Model, and precise measurements of the
inconsistencies could
give some clues to the flavor and CP structure of the new physics.
The \pnn~ decay modes are particularly attractive due to the fact that
the meson decays can be cleanly related to
the short-distance Feynman diagrams at the quark level, allowing a
precise extraction of the quark mixing parameters.  The theoretical
uncertainties in the branching ratios are estimated to be $\sim$7\% for
\kpnn~ and $\sim$2\% for \klpnn~ with an anticipation that the
uncertainty for the charged mode could be brought down to $\sim$2\%
with a NNLO QCD calculation \cite{Buras04}.  Promising quantities for
comparison include: 1) a comparison of BR(\kpnn) with
the ratio $\Delta m_d/\Delta m_s$ from $B_{d,s}$ mixing,
2) a comparison of the ratio BR(\klpnn)/BR(\kpnn) with $\sin2\beta$
from $B_d$ decays,  and 3) a
comparison of the Jarlskog invariant \cite{Jarlskog}, proportional to
the area of the unitarity triangle, between K and B decays.
Experimental challenges include the poor kinematic signature due to
the 3-body final state with two neutrinos and the low branching ratio.
Current estimates for the charged and neutral modes are $0.78 \pm 0.12
\times 10^{-10}$ and $0.30 \pm 0.06 \times 10^{-10}$,
respectively \cite{Buras04},
where the quoted uncertainties are dominated by the measurement
uncertainties on the parameters of the CKM matrix.

\section{E949: \kpnn}

Following the observation by E787 of two clean candidates of the
decay \kpnn, the E949 experiment was proposed in 1998 to improve the
sensitivity by about one order of magnitude.  First data were taken in
2002 and first results have been published recently \cite{E949pnn1}.

E949 is based on ``modest'' upgrades to the E787 apparatus; as such,
the background level and signal sensitivity can be predicted with confidence.
The experiment uses the entire proton flux from the AGS, increasing the
proton intensity from $15\times 10^{12}$ per spill to
$65\times 10^{12}$; however, to keep instantaneous rates constant in
the detector, the spill length is increased accordingly so we do not
gain linearly with proton intensity.
The running time per year is increased to $\sim
25$ weeks, taking advantage of the long RHIC running.  Detector
upgrades included an addition to the barrel photon veto to increase
the number of radiation lengths, better photon veto coverage along the
beam direction, higher segmentation of beam tracking
elements, improved pion tracking resolution via changes to the
readout electronics of the central drift chamber and range stack straw
chambers, improved pion energy resolution by replacing
scintillator and by the addition of a phototube gain monitoring
system, and finally, upgrades to the trigger/DAQ to handle higher data rates.

\subsection{Detector}

\begin{figure}[htb]
\begin{center}
\epsfig{file=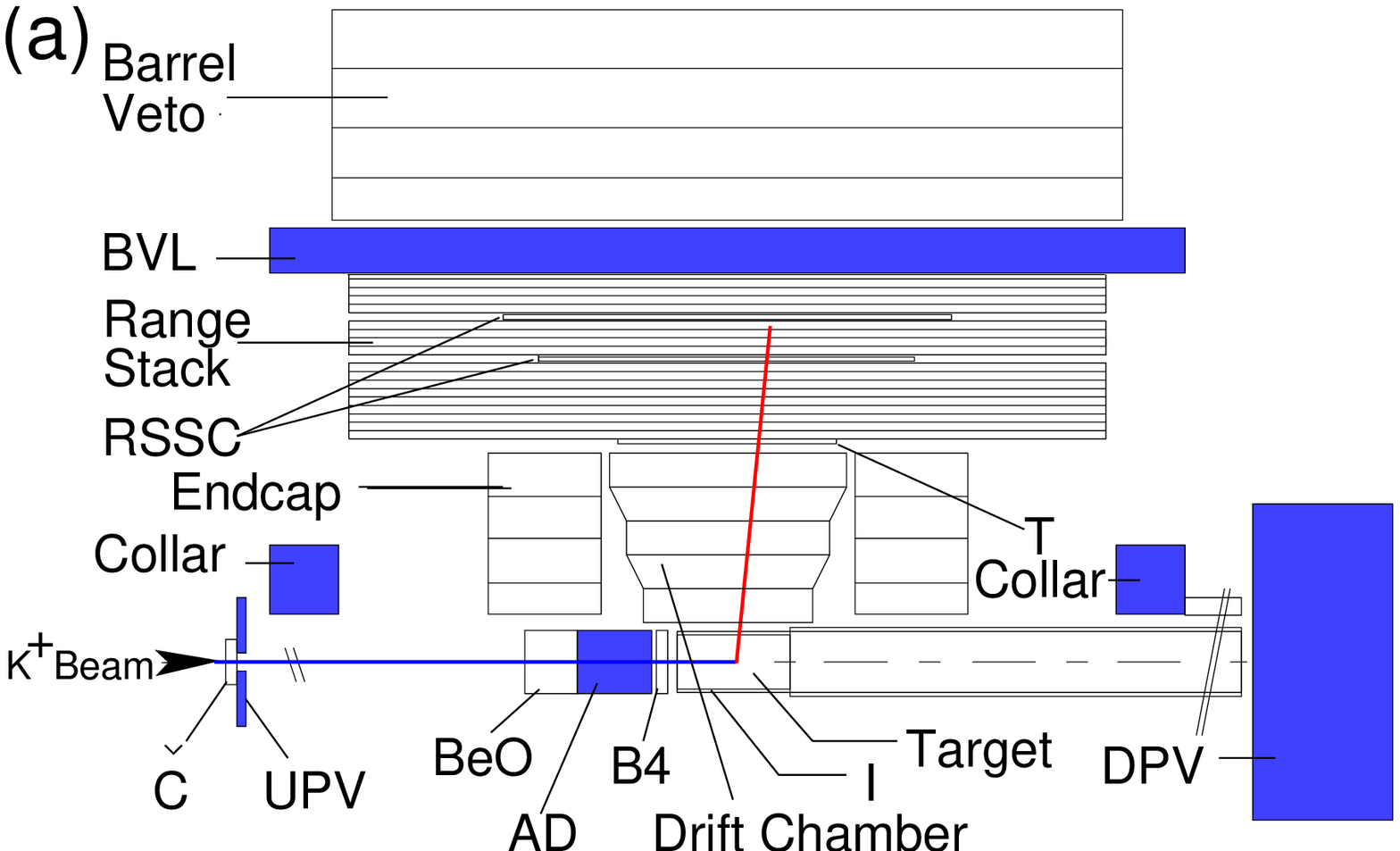,width=16pc}
\vspace{0.1cm}
\epsfig{file=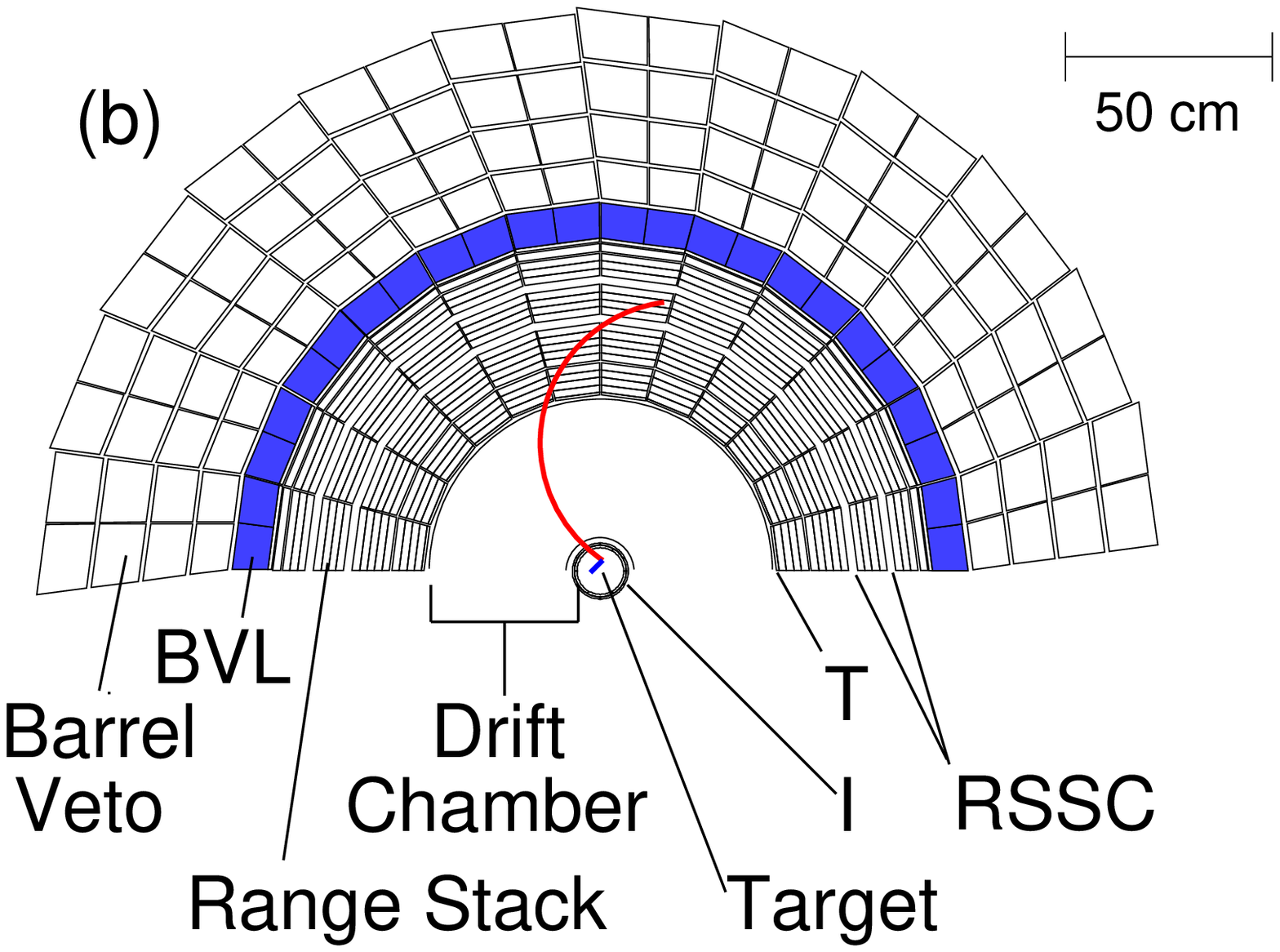,width=16pc}
\caption{Side (a) and end (b) views of the E949 detector.  The
  detector has cylindrical symmetry; only the top half is shown.
  Additions to the photon veto system for E949 are shown in blue.}
\label{fig:E949det}
\end{center}
\end{figure}

The E949 detector is shown in Figure \ref{fig:E949det}; most details
can be found from the references in \cite{E949pnn1}.

The incoming $K^+$ beam ($K/\pi$ ratio ideally about 4) with a
momentum of $\sim 700$ MeV/c is tracked by Cerenkov counters, MWPCs and
scintillator hodoscopes, slowed down in a BeO degrader, and comes to
rest in a scintillating fiber target.  The stopped kaon technique
allows for large geometric acceptance and 
causes the dominant backgrounds of \kpitwo~ and \kmutwo~ to appear
kinematically as monochromatic peaks, as well as suppressing scattered
beam pions.  The K decay products are momentum analyzed in a 1T
cylindrical field and then stopped in a 
scintillator ``range stack'', allowing measurements of kinetic energy
and range to provide redundancy in the kinematics.  The \pme~ decay
sequence of the
stopped pion is observed by waveform digitizing the scintillator
signals, providing a powerful handle against \kmutwo~ background.
Photon detectors surround everything and inactive material in the
detector is minimized.

\subsection{2002 run}

The proton intensity during the 2002 run was according to design,
although it must be said that we did not run concurrently with RHIC.
The
typical intensity was
$65\times 10^{12}$
protons in a 2.2 sec spill with a 3.2 sec interspill.  Peak
intensity reached $76\times 10^{12}$ protons/spill.  However, there
were several non-optimal features of the 2002 run, the first being the
short duration, about 12 weeks in total.  Next, damage to the primary AGS
motor generator set meant running with the backup supply, forcing us to lower
the proton momentum to 21.5 GeV with a resulting $\sim 10$\% loss in
K flux; this also
resulted in
20\% (40\%) worse duty factor compared to E787 (E949
proposal).
Finally, problems with the
beamline separators resulted in a $K:\pi$ ratio of about 2, compared
to a typical value of 4 in E787; the resultant instantaneous rates were roughly
twice those seen in E787. The integrated kaon flux was about one-third
of the total accumulated by E787 over $\sim80$ weeks.

\subsection{Backgrounds}

The main sources of background are the two-body decays \kmutwo~ (\km)
and
\kpitwo (\kp), multibody decays containing muons (\kmm) such as
$K^+\rightarrow \mu^+ \nu \gamma$, $K^+ \rightarrow \mu^+ \pi^0 \nu$,
and \kp~ with $\pi$ decay-in-flight, scattered beam pions (either from kaon decay/interaction or
pions from the $K^+$ production target) and $K^+$ charge exchange (CEX)
reactions, resulting in decays $K^{0}_{L} \rightarrow \pi^+ l^-
\overline\nu$, where $l=e$ or $\mu$.
To elude rejection, \km~ and \kp~ events have to
be reconstructed incorrectly in range, energy and momentum.  In
addition, any event with a muon has to have its track misidentified
as  a pion.  The most effective weapon here is the waveform digitizer analysis,
requiring observation of the \pme decay sequence; this provides a
muon rejection factor of about $10^{5}$.  Events with photons, such
as  \kp~ decays, are efficiently eliminated by the photon veto; the
rejection factor for events with $\pi^{0}s$ is around $10^{6}$.  A
scattered beam pion can survive the analysis only by
misidentification as a $K^+$ and if the track is mismeasured as
delayed, or if the track is missed entirely by the beam counters
after  a valid $K^+$ stopped in the target.  CEX background events
can  survive only if the $K_{L}^{0}$ is produced at low enough
energy  to remain in the target for at least 2 ns, if there is no
visible gap between the beam track and the observed $\pi^+$ track,
and  if the additional charged lepton goes unobserved.

\subsection{Offline analysis}

The heart of the offline analysis is a robust estimate of the
background at the level of a fraction of an event.  The key features
of the analysis strategy are summarized below.

The sources of background are identified {\it a priori},
as listed above.  The question of backgrounds that
may not have been identified in advance will be addressed briefly later.

The same dataset is used for the background studies and signal
search. This ensures that the impact on the background estimates
of time-dependent effects,
intermittent effects (like hardware failures) or intensity-dependent
effects
are properly taken into account.

The analysis is
done blind, meaning that the signal region is hidden (by inverting
cuts) while cuts are developed and background levels estimated.  The
background levels themselves are also estimated in a blind way in order
to eliminate bias from cut tuning on the relatively small number
of events left at the end of a typical background study.  A uniformly
selected subset of the data is used to develop the cuts
against the background;
the effect of these cuts is then measured (once) in an
unbiased way on the remainder of the data.

Two independent cut sets, each with high rejection, are used for
each background.  These cut sets are played off against each other to
measure their background rejection power.  For example, for \kp~
background, the two independent cut sets are the kinematic cuts and the
photon veto cuts. By inverting the photon veto cuts (i.e. by demanding
the presence of photons), after first removing the non-\kp~ backgrounds,
one can isolate a sample of \kp~ events
from which the kinematic rejection can be measured.  And vice versa.
The rejection of the two cut sets can be multiplied to give the total
rejection if the cuts are not correlated.
Ideally the rejections of the two cut sets are comparable so that the
rejection of each
cut set can be studied with adequate statistics.
For the \km~ and \kmm~ backgrounds, the kinematic cuts are played off against
the waveform digitizer cuts.  For scattered pions, the delayed
coincidence cuts (cuts
that ensure that the kaon stopped, then decayed after a suitable time
interval) are played off against particle ID cuts in the beamline hodoscopes.
For the case where the kaon and pion are from different particles, the
delayed coincidence cut can be fooled; for this class of events, cuts
on the pattern of hits in the kaon stopping target are played off
against cuts that detect multiple tracks in the beamline wire chambers
and hodoscopes.  The exception to this methodology
is the CEX background which is
determined largely by Monte Carlo.

Correlations between the independent cuts can spoil the background
estimation.  We look for explicit evidence for correlations by looking
at variations in the rejection of each independent cut as a function
of a large set of variables and then seeing if the partner cut shares
similar dependences.  Problematic regions in cut space are removed by
so-called ``setup'' cuts (more on these in a moment).  As a test of
the method, the numbers of observed events in regions near the signal
region are compared with the predicted background rates based on the
product of the rejections of the independent cut set pairs.  The
observed consistency gives us confidence that the cuts are independent
into the signal region. We also
examine all the events that fail only the ``setup'' cuts to make sure that
they are tight enough to remove all correlations; this is also an
opportunity to search for hitherto unknown backgrounds since the
primary backgrounds have been largely suppressed.

The assessment of candidate events and the computation of the \kpnn~
branching ratio are done with a likelihood ratio technique \cite{Junk}.
An event characterization function consisting of a set of discrete
bins $S_i/b_i$ describes the relative probability for events occurring
in bin $i$ to originate from \kpnn~ or background. Here $b_i$ is the
expected number of background events from all sources in bin $i$, and
$S_i$ is the expected number of signal events in each bin, given as 
$\mathcal{B}A_{i}N_{K}$ where $A_i$ is the acceptance in bin $i$ and
$\mathcal{B}$ is the branching ratio from the fit.
Rather than trying to explain the details, let me just state that the
validity of the technique was studied extensively and confirmed
with Monte Carlo experiments.  Reflecting a growing confidence in
our ability to predict the background level and a shift in philosophy
from ``discovery'' mode to ``measurement'' mode, we expanded the
E949 pre-determined signal region, letting in more
background but also allowing us to gain back about 30\% in acceptance.
The total background expected in the signal region was $0.30 \pm
0.03$ events, dominated by \kp~ background ($0.216 \pm 0.023$).  The \km, \kmm,
and beam-related backgrounds (including CEX) contributed
$0.044\pm0.010$, $0.024\pm0.003$, and $0.014\pm0.003$ events, respectively.

\begin{figure}[t]
\begin{center}
\epsfig{file=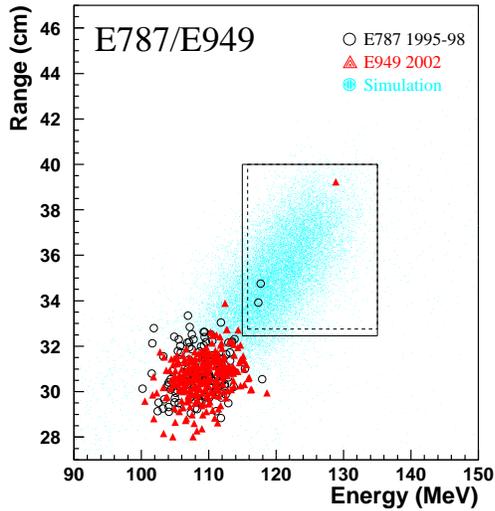,width=17pc}
\caption{Range (R) vs. energy (E) distribution of events passing all
  other cuts of the final sample.  The circles represent E787 data and
  the triangles E949 data.  The group of events around E=108 MeV is
  due to $K_{\pi2}$ background.  The distribution of \kpnn~ events
  from Monte Carlo is shown by dots.  The solid (dashed) line box
  represents the signal region for E949 (E787).}
\label{fig:E949result}
\end{center}
\end{figure}

After all cuts were applied, one  candidate near the \kpnn~
endpoint was observed, as shown in
Figure \ref{fig:E949result} (together with the previous data from
E787).
The estimated probability that the background alone gave rise to this
event (or any more signal-like event) was 0.07.  At the measured
branching ratio (see below), the $S_i/b_i$ for
this event was 0.9, compared to values of 50 and 7 for the previous
two candidate events seen by E787.  The best estimate of the branching
ratio, combining data from E787 and E949 is $\mathcal{B}$(\kpnn)$ =
1.47^{+1.30}_{-0.89}\times 10^{-10}$, consistent with Standard Model
expectations, but intriguing enough to justify completion of the
experiment.

\section{KOPIO: \klpnn}

The best direct experimental limit on \klpnn~ comes
from KTeV \cite{KTeV97} where 
the Dalitz decay of the $\pi^0$ was used to locate the $K_L$
decay vertex, reaching a limit $\mathcal{B}$(\klpnn) $< 5.9 \times
10^{-7}$ (90\% CL).  A model-independent bound (``Grossman-Nir'' bound)
can be derived \cite{Grossman-Nir} from $\mathcal{B}$(\kpnn); with the
latest numbers from E949, this leads to a better limit of
$\mathcal{B}$(\klpnn) $< 1.4
\times 10^{-9}$ (90\% CL).   Future searches will all utilize the $\pi^0
\rightarrow \gamma\gamma$ mode.  A test of the technique with a
narrowly collimated high energy beam (so-called ``pencil beam'') was
performed by KTeV \cite{KTeV96}.  E391a at KEK is the first dedicated
experiment to search for \klpnn~ using a pencil beam; data-taking
took place this year from February
through June.  If they can get away with very loose photon veto
cuts, they might be able to improve on the Grossman-Nir
bound from E949 \cite{KomaDafne}.

KOPIO takes a very different approach.  A low energy $K_L$ beam
(momentum around 800 MeV/c) is obtained at a
$45^{\circ}$ production angle.  The large angle suppresses hyperon
production and the soft neutron spectrum reduces the production of
$\pi^0$s from neutron interactions. The ``pancake'' beam (5mrad vertical
vs 100 mrad horizontal) arrives in 200ps-wide
microbunches every 40ns, allowing time-of-flight to be used to determine the
$K_L$ momentum.  The interbunch extinction is expected
to be $\sim 10^{-3}$.
KOPIO will run with $100 \times 10^{12}$ protons on
target; this requires an
AGS injector upgrade.  About $3\times10^{8}~ K_L$ are produced per
spill, of which about 12\% decay in the fiducial volume;
these are accompanied by about $3\times 10^{10}$ neutrons.

\begin{figure}[htb]
\begin{center}
\epsfig{file=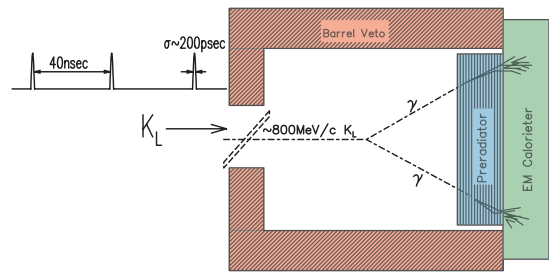,width=17pc}
\caption{Concept of the KOPIO detector.}
\label{fig:kopio_concept}
\end{center}
\end{figure}

The concept is shown in Figure \ref{fig:kopio_concept}.
$K_L$ decays are observed in a 4m-long vacuum tank ($\sim 10^{-7}$
Torr) surrounded by scintillator sheets for charged particle vetoing,
followed by photon vetoes, using Shashlyk
technology \cite{Atoyan92} in the barrel region and
Pb-scintillator ``logs'' in the
upstream wall. The direction of forward photons
is measured in a
preradiator consisting of alternating layers of radiator, wire
chambers, and scintillator.
 Together with a constraint from the
flat beam, this allows reconstruction of the $K_L$ decay vertex.
Photon energies are measured primarily in a Shashlyk
calorimeter.   The full kinematics of the $K_L$ decay can therefore
be determined, suppressing many backgrounds, and serving to relax
the photon veto requirements. Furthermore, this provides a second
independent cut set (along with the photon or charged veto) and
allows a powerful
technique for measuring the backgrounds as demonstrated by E787/E949.

\subsection{Background suppression}

\begin{figure}[htb]
\begin{center}
\epsfig{file=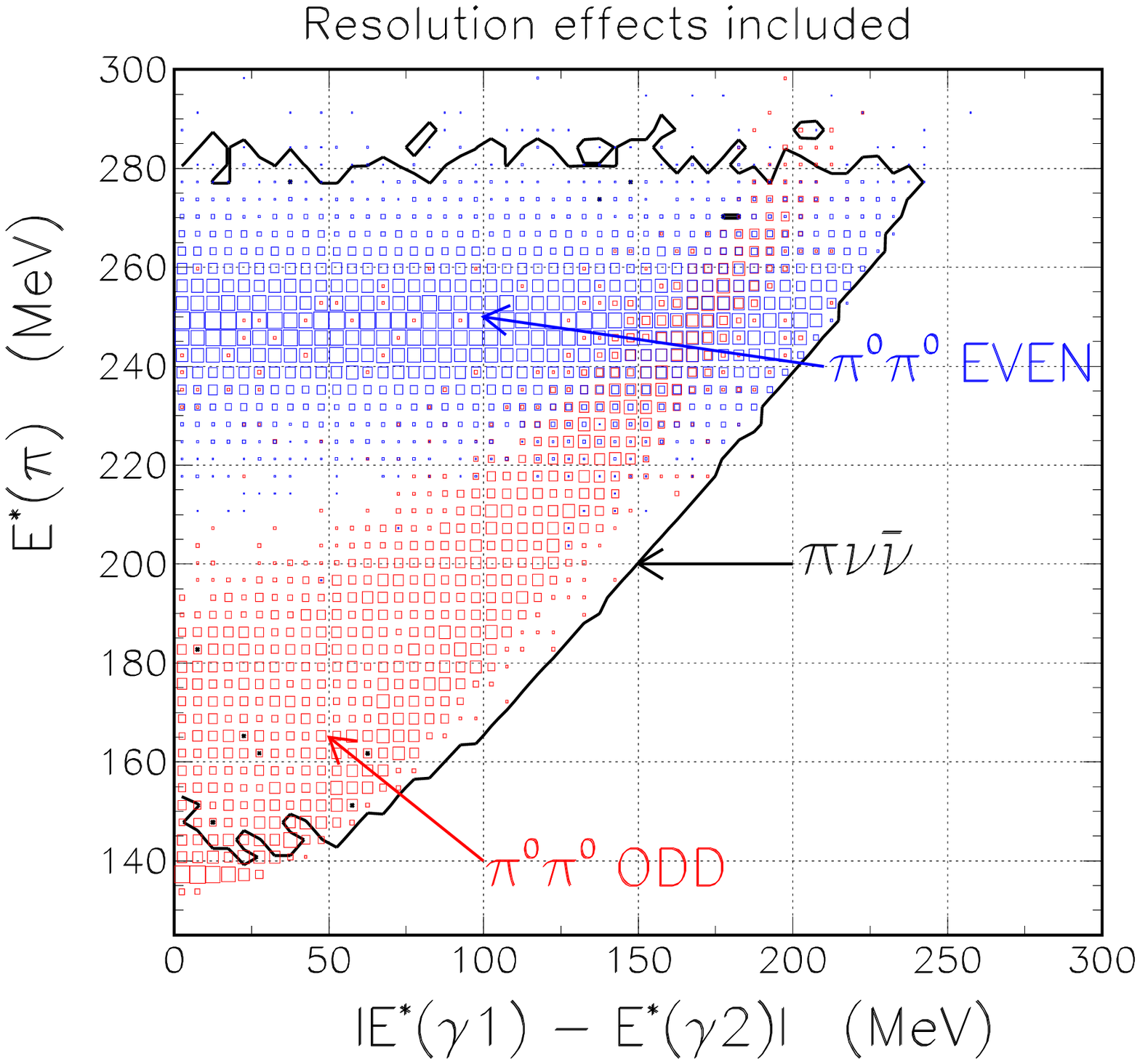,width=16pc}
\caption{\klpitwo~ background.  $\pi^0$ energy versus the
  difference in energy of the two detected photons, measured in the
  $K_L$ rest frame.  The events are shown after a cut on the two
  photon invariant mass, which constrains the ``odd'' background  to
  lie on a band.}
\label{fig:kpi2kin}
\end{center}
\end{figure}

The dominant background is \klpitwo~ with two missed photons.  These
can be divided into two topologies, one where both photons from one
$\pi^0$ are missed (``even'') and the other where one photon from
each $\pi^0$ is missed (``odd'').  The power of the kinematic
rejection is shown in Figure \ref{fig:kpi2kin}.  Cuts on the $\pi^0$
energy (\epistar) and on the difference in energy
(\ediff) between the
two photons (all measured in the $K_L$ rest frame) are effective.

To fully suppress \kp~ background, a $\pi^0$ detection inefficiency of
$10^{-8}$ is required.  E787 obtained a $\pi^0$ rejection of
$10^{-6}$ for 200 MeV/c $\pi^0$s which yield photons between 20 and
225 MeV. Inefficiency for single photon detection ranged from about
$10^{-2}$ at 20 MeV down to about $10^{-4}$ at around 200 MeV. These
measurements are currently being redone with the E949 apparatus.  To
reach $10^{-8}$ rejection in KOPIO, we take advantage of kinematic
handles to reject events with low energy missing photons.  The
energy of the missing photons can be obtained by
subtracting the measured energies of the two observed photons from the
$K_L$ energy; requiring significant missing energy suppresses events
containing lower energy missing photons.  For asymmetric $\pi^0$
decays, a cut on the missing mass is effective since the missing mass
is proportional to $\sqrt{E_{miss\gamma 1} \cdot E_{miss\gamma 2}}$.
The KOPIO goal is to obtain a photon detection inefficiency that is a
factor of 3 lower than the E787 measurements; this is thought to be
achievable by going to finer sampling (more radiation lengths) for
lower (higher) energy photons.  Photons escaping through the downstream
beam hole are detected by lead-aerogel Cerenkov counters placed in the
downstream section of the beam.
The number of photons escaping through the upstream
beam hole was found to be negligible.

The \kp~ ``even'' background arising from a slow $K_L$ (or neutron)
from the previous
microbunch can be dealt with by looking at the correlation between
\epistar~ and the longitudinal $\pi^0$ momentum, where \epistar~ is
calculated assuming that the particle came from the previous
microbunch.  The background is then cleanly localized whereas the
signal is spread out.

\begin{figure}[htb]
\begin{center}
\epsfig{file=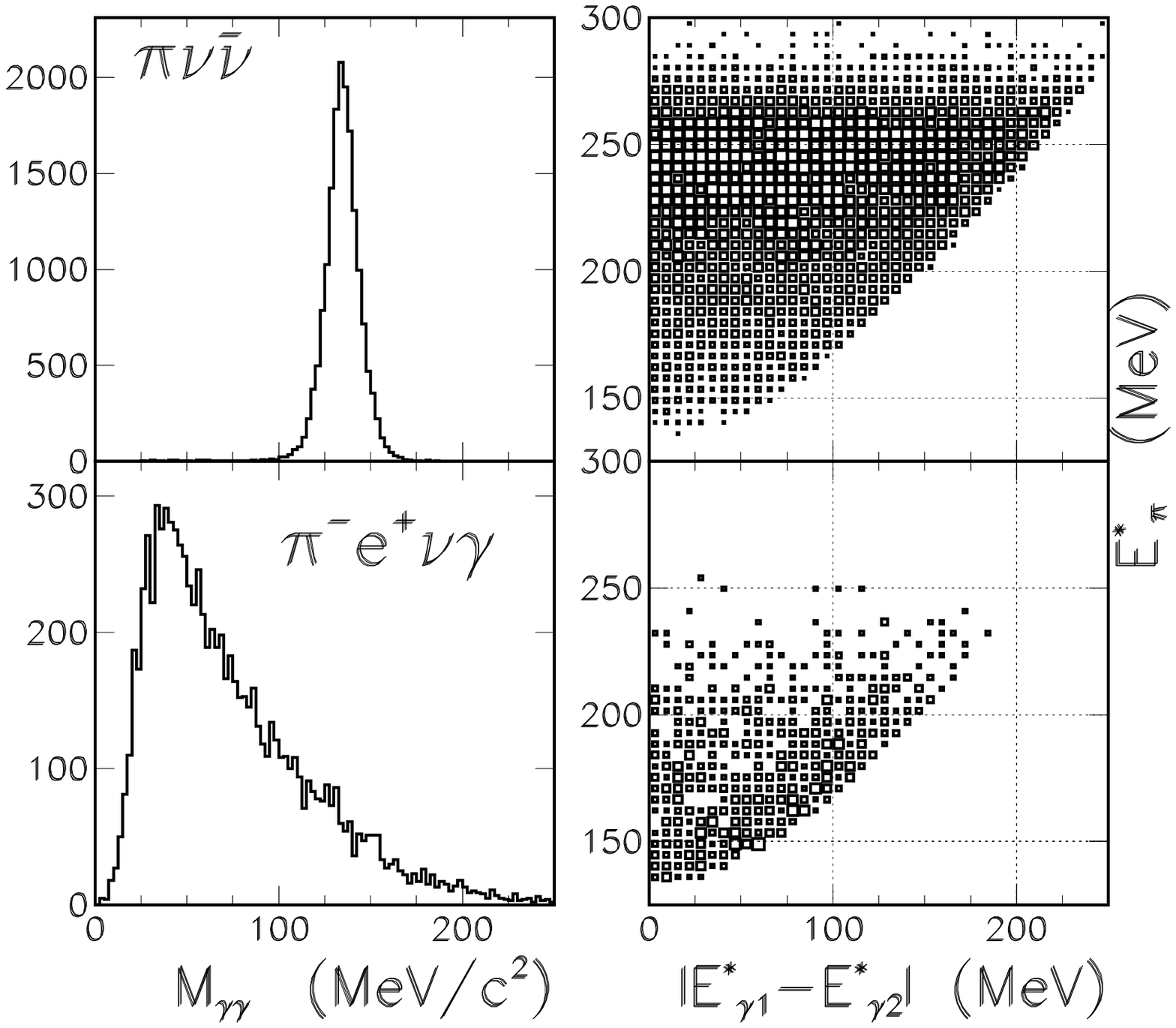,width=16pc}
\caption{\klethreeg~ background.  The plots on the left show the two
photon invariant mass for signal (top) and background (bottom).
The plots on the
right show the plane \epistar~ vs. \ediff~ for signal and background.}
\label{fig:ke3gkin}
\end{center}
\end{figure}

To suppress backgrounds containing charged particles, detection
inefficiencies for $e^-$,$e^+$,$\pi^-$,$\pi^+$ of better than 
$10^{-5}$,$10^{-4}$,$10^{-4}$,$10^{-5}$ are required; this seems achievable
based on beam tests.
The dominant charged-mode
background is \klethreeg~ where the positron converts asymmetrically
before detection in the charged veto, the low energy $\gamma$ from the
conversion is missed, and the $\pi^-$ is missed.  Cuts on the two
photon mass as well as on \epistar~ and \ediff~ are effective here as
shown in Figure \ref{fig:ke3gkin}.

Hyperon decays (primarily $\Lambda$) are mostly suppressed by the large angle
neutral beam; the production cross section is low and the momentum is
low so that all $\Lambda$s decay before reaching the fiducial volume.
The production of $\Lambda$s by interactions of the neutral beam with
collimators is reduced by the low momentum; good collimation of the
beam and good vertexing ensure that events produced near the last
collimator are not a problem.
The production of $\pi^0$s from neutron interactions is largely suppressed by
having excellent vacuum.  The large beam angle also reduces the number
of neutrons above $\pi^0$ production threshold.  The kinematic cuts
for \kp~ background further suppress the neutrons due to the
misassignment of a neutron for an incoming $K_L$.

A summary of the expected signal and background for the projected
4-year running period of KOPIO is shown in Table \ref{tab:sb}.
Assuming the Standard Model branching ratio, about
40 signal events are expected over a background of 20, leading to a
20\% measurement of the branching ratio, or a 10\% measurement of
Im($\lambda_t$).

\begin{table}[htb]
\caption{Expected number of signal and background events assuming the
Standard Model branching ratio and 3 years of KOPIO running.}
\label{tab:sb}
\begin{tabular}{@{}lc}
\hline
Process & Events \\
\hline
\klpnn & 40 \\
\hline
\klpitwo & 12.4 \\
$K_L \rightarrow \pi^{\pm} e^{\mp} \nu \gamma$ & 4.5 \\
$K_L \rightarrow \pi^+ \pi^- \pi^0$  & 1.7 \\
$K_L \rightarrow \pi^{\pm} e^{\mp} \nu$ & 0.02 \\
$K_L \rightarrow \gamma \gamma$ & 0.02 \\
$\Lambda \rightarrow \pi^0 n$ & 0.01 \\
Interactions ($nN \rightarrow \pi^0 X$) & 0.2 \\
Accidentals & 0.6 \\
\hline
Total background & 19.5 \\
\hline
\end{tabular}
\end{table}

\subsection{Recent developments}

Recent work has concentrated on firming up the detector design.

Studies of the beam include tests of the microbunching width and the
interbunch extinction.  Measurements in a test run with a 93 MHz RF
cavity showed a 240ps width, compared to about 215ps expected from
simulation.  Using the same simulation to extrapolate to KOPIO running
conditions (25 MHz cavity to get the 40ns microbunch spacing and a 100
MHz cavity to get the microbunch width), yields a microbunch width of
185ps so this seems to be in good shape.  For the interbunch
extinction, a level of about $10^{-3}$ is needed; the test run with
the 93 MHz cavity yielded an extinction of about 0.015.  A new test
run with a 4.5 MHz cavity was just completed this past June.
A $\overline{p}$ beam was used
to improve the systematics (previously a photon beam was used).  Bunch
width and extinction measurements were made in a matrix of RF
frequency, RF voltage and $\Delta p/p$; offline analysis and 
comparisons to simulation are in progress.

\begin{figure}[t]
\begin{center}
\epsfig{file=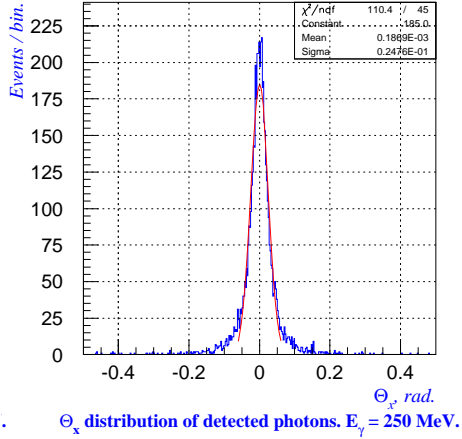,width=16pc}
\caption{Angular resolution of a preradiator prototype for 250 MeV
  photons.  The fitted sigma is 25 mrad.}
\label{fig:PR_angle}
\end{center}
\end{figure}

The preradiator consists of four $3.5 \times 3.5$m$^2$ quadrants
(active region $2 \times 2$m$^2$).  Each quadrant contains
64 layers, where each layer is composed of a sheet of extruded
scintillator with wavelength shifting (WLS) fiber readout,
wire chamber with cathode strip
readout and radiator (copper and aluminum); the thickness of
each layer is about $0.03 X_0$. 
The preradiator concept has been tested in a tagged photon beam at
the LEGS facility at BNL.  The angular resolution for 250 MeV photons
is shown in Figure \ref{fig:PR_angle}. The fitted sigma is 25 mrad.
Currently in progress are full-scale prototyping, high voltage and
readout electronics, QA studies of the scintillator production facility,
and a full mechanical design.

\begin{figure*}[htp]
\begin{center}
\epsfig{file=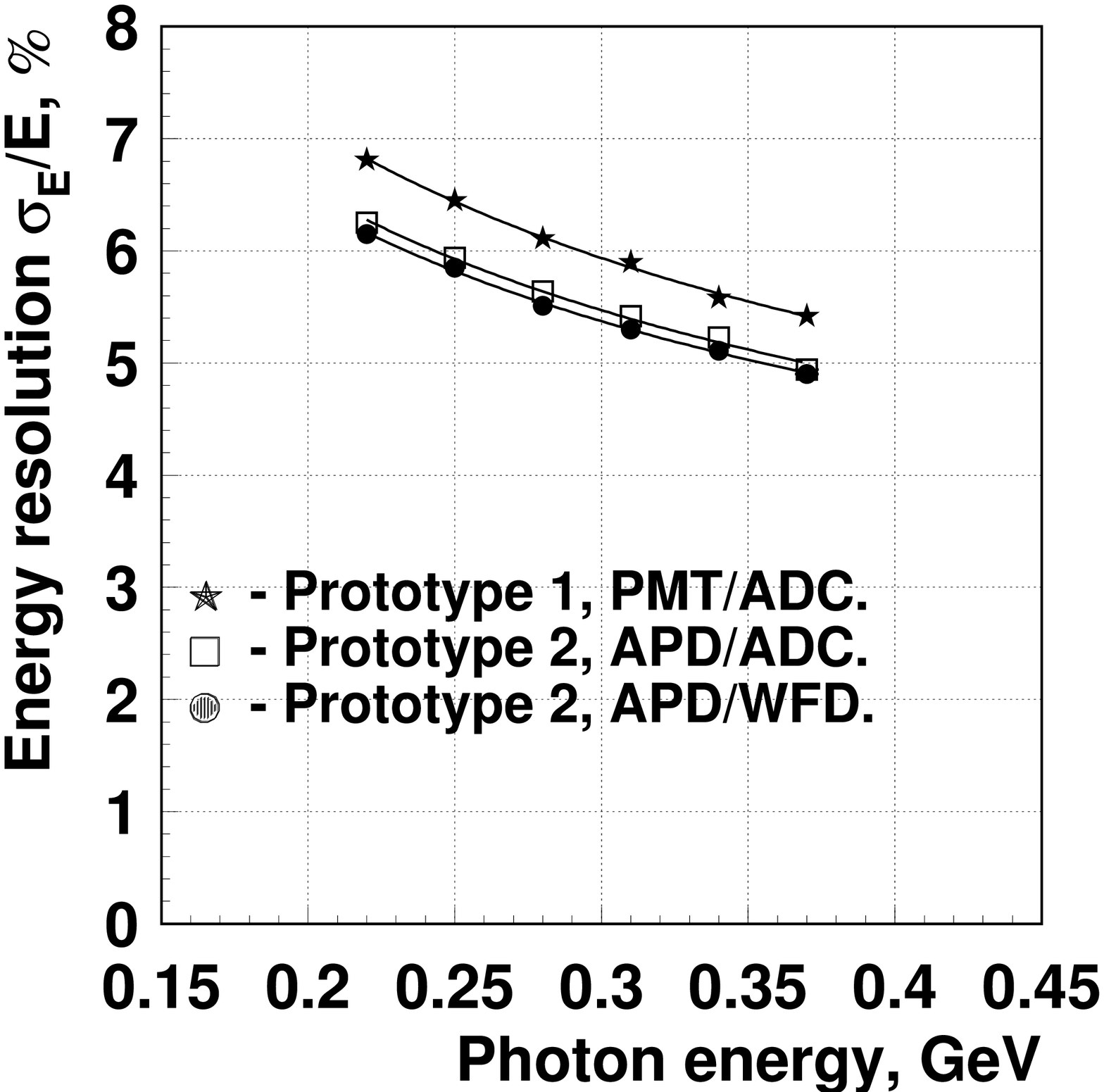,width=13pc}
\epsfig{file=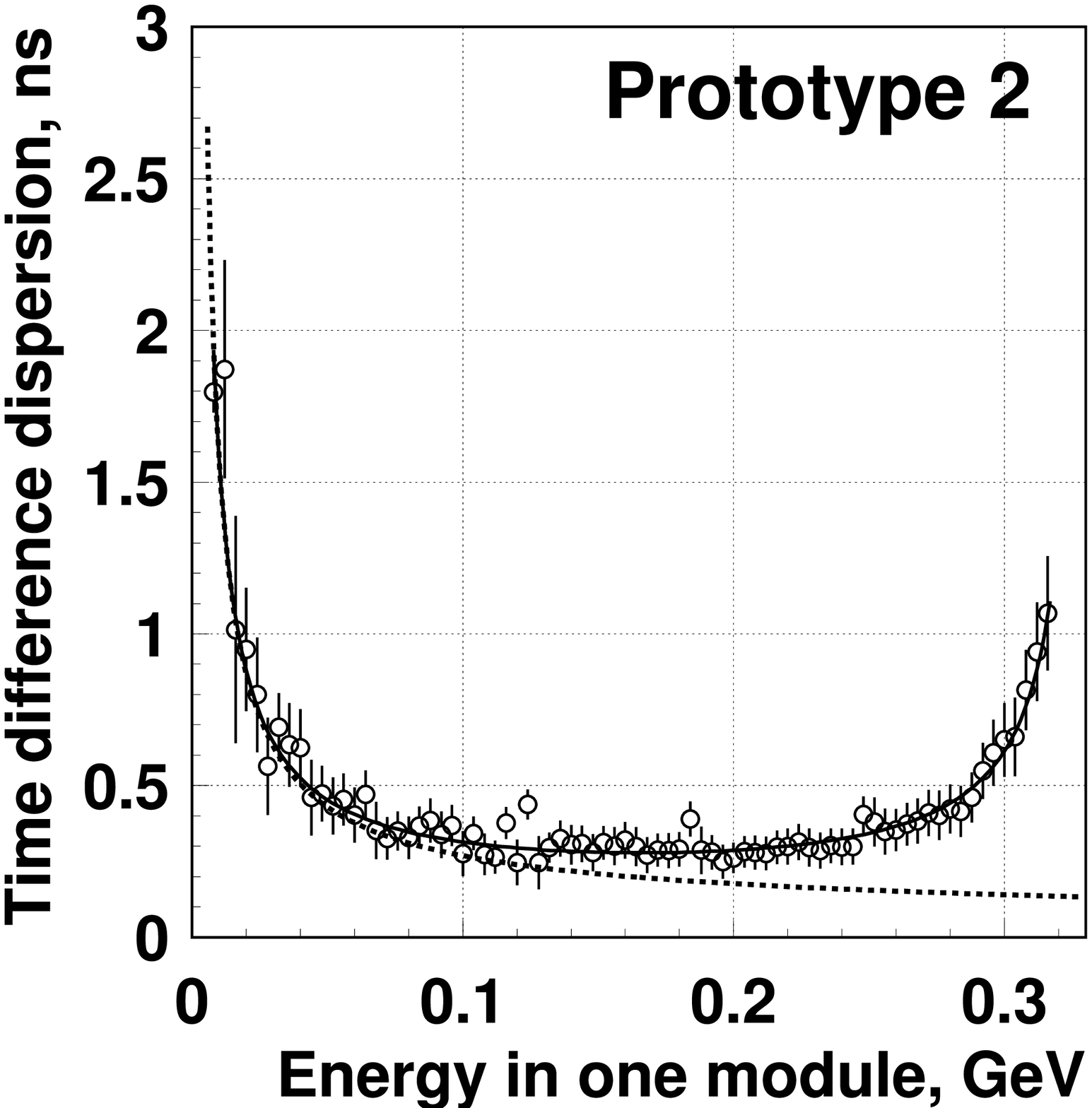,width=13pc}
\caption{Left: test beam measurements of the energy resolution versus
  photon energy of the
  prototype Shashlyk calorimeter.  Right: rms of the time difference
  between two Shashlyk modules versus the energy in one module.  The
  rms rises above 0.3 GeV because there is very little energy
  deposited in the second module.  The time resolution in a single
  module is shown by the dotted line. From \cite{Atoian04}.}
\label{fig:Shashlyk}
\end{center}
\end{figure*}

The Shashlyk 
calorimeter has been
extensively studied \cite{Atoian04}. It
consists of a 48 $\times$ 48 array of 
modules where each module is 11cm $\times$ 11cm $\times$ 65cm in
depth, consisting of 300 alternating layers of 0.275mm
thick lead and 1.5mm thick scintillator (16 $X_0$).  The
WLS
fibers are read out with an avalanche photodiode (APD).
 Test beam results for energy and
timing resolution are shown in Figure \ref{fig:Shashlyk}.
A fit to the energy resolution yields $\sigma_{E}/E = (2.9 \pm
0.1)\%/\sqrt{E(GeV)}$.   The timing resolution was obtained by looking
at the time difference between two Shashlyk modules as a function of
the energy in one of the modules.  A fit yields $\sigma_{t} = (90 \pm
10)ps/\sqrt{E(GeV)}$. Both energy and timing resolution are within the
KOPIO specification, and effort has moved on to the full mechanical
design, high voltage and readout electronics, and systems for quality
control, gain monitoring and cooling of the APDs.

The photon veto system in the barrel region will use esssentially
the same Shashlyk technology as in the calorimeter, but with
0.5mm/1.5mm lead/scintillator layers, and a total thickness of 17
$X_0$.
Studies are ongoing to optimize the photon detection efficiency, to
understand the mechanical integration with the vacuum tank and to
increasing the signal acceptance by taking events where one photon
converts in the barrel veto.
The upstream wall will consist of ``logs'' of 1mm/7mm
lead/scintillator layers read out with WLS fibers \cite{Mineev02}.
Light yield,
timing and long-term stability tests have been performed on prototypes.
Extrapolating to the final design, timing and energy resolutions of $\sim
70ps/\sqrt{E(GeV)}$ and $5-6\%/\sqrt{E(GeV)}$, respectively are expected.

For the charged veto system, the response of plastic scintillator to
$\pi^{\pm}$, $\mu^{\pm}$~ and $e^{\pm}$ has been measured in the
momentum range of 185-360 MeV/c in a test
beam at PSI and compared to simulation.  In order to reach a $\pi^-$
detection inefficiency of better than
$10^{-4}$ in the charged veto system alone, the $\pi^-$ must be
detected before it traverses more than
0.3mm in scintillator and undergoes charge exchange into neutral
products; this
corresponds to a detection threshold of
$\sim 75$ keV.  In reality, some of the neutral products
(e.g. photons) will be detected in the photon veto system.
In addition, the amount of dead material in front of
the veto system must be kept below 20 mg/cm$^2$; this puts stringent
requirements on the membrane separating the high vacuum of the decay  
volume from the lower vacuum where the charged vetoes will be
situated.
Current design for the charged veto system in the vacuum tank envisages
a single layer of 2mm-thick overlapping sheets of
plastic scintillator with direct phototube
readout.  Tests showed that this approach allows a detection
threshold as low as 10 keV and $\pm5$\% light collection uniformity
over the detector surface.  Studies are ongoing on optimizing the
thickness, light output, and reflector material as well as on
mechanical and vacuum-related
issues. In addition, the interplay between the charged and photon vetoes
for those cases where the charged particle converts into photons is
being studied to optimize the performance of the overall system.

To detect photons passing through the beam hole, an array of
lead-aerogel counters (collectively called ``catcher'') will be
deployed in the beam.  The current
design calls for a module of 2mm lead, followed by 5cm of aerogel
(n=1.05) with a flat mirror directing the Cerenkov radiation from the
photon conversion to a 5-inch PMT.  Around 400 modules will be
placed in 25 layers along the beam direction for a thickness of $8.3
X_0$ in total.  Prototypes have been tested for light yield and for
the response to protons (as a substitute for neutrons) and good
agreement with simulation was seen.  Extrapolating to the final design,
a photon efficiency $>99$\% for energy 
above 300 MeV and a neutron sensitivity of 0.3\% for energy around
800 MeV (typical for our beamline) is expected.  To cover shower
escape from the lateral edges of the catcher for those photons
entering at very oblique angles, an array of 2mm/10mm
lead-plastic Cerenkov counters (``guard'' counter)
will be placed in the neutron halo
region.  Current efforts are directed towards testing of full-scale
prototypes, studies of veto blindness by photons (from the production
target), neutrons, and $K_L$ decays in the catcher,  and estimating
rates in the guard counters.

\section{Outlook}

To clarify the current experimental situation on \kpnn, it would be
desirable to complete the E949
program (60 weeks of running in total), but unfortunately AGS
operations for HEP were
cancelled after the 2002 run. In the meantime, a proposal to complete
E949 has been submitted to the NSF.
There is good reason to believe that E949 can achieve its design
goal if it were to run for the designed length of time;
the 2002 run showed that the
E949 upgrades worked and that the backgrounds are well understood.
One might even speculate, based
on our experience with the higher rates in 2002, 
that it would be possible to take more instantaneous beam
in the future, allowing us to take better advantage of the higher
proton intensity being delivered by the AGS.  There is also the
potential of increasing the sensitivity in the kinematic region below
the \kp~ peak; this region was background limited in E787
analyses \cite{pnn2_96}, including the recently published result on the
1997 data \cite{pnn2_97}, but it is hoped that improvements made in E949 to
the photon veto system will make this kinematic region viable.
Analysis of the 2002 data in this region is currently in progress.
Photon veto improvements in the barrel region have already been verified in the
E949 analysis above the \kp~ peak; what remains to be seen is the (crucial)
photon coverage in the beam direction.

While BNL remains firmly committed to E949, 
experiments have been proposed at other labs to take the sensitivity
one step further to the level of between 50 and 100 events.  These
include decay-in-flight experiments P940 at
Fermilab \cite{Cooper},
NA48/3 at CERN \cite{Ceccucci} and a stopped kaon experiment at
JPARC \cite{Komatsubara}.

With regards to \klpnn, the detector R\&D phase of KOPIO is starting
to wind down.  Key
features of the concept have been established, and planning for the
construction phase is beginning.  The KOPIO project is part of a
larger NSF project (``Rare Symmetry Violating Processes'', or RSVP)
that includes a search for muon-electron conversion (MECO \cite{Meco})
at the BNL AGS.  RSVP was included in the FY05 President's Budget
for a construction start in 2005; a 5-year construction is
envisaged at a cost of about \$140 million, with operations starting
in 2008. 

\section{Acknowledgements}

Many thanks to the organizers for a most enjoyable workshop in
an incomparable setting.  Thanks also to A. Ceccucci for the
invitation, and to L. Littenberg and S. Kettell for a careful
reading of the manuscript. This work was supported in part under US
Department of Energy contract DE-AC02-98CH10886.

\end{document}